\documentclass[letterpaper, 10 pt, conference]{ieeeconf}  

\IEEEoverridecommandlockouts                              
\title{\LARGE \bf
Stochastic Games for Security in Networks with Interdependent Nodes
}

\author{Kien C. Nguyen, Tansu Alpcan, and Tamer Ba\c{s}ar
\thanks{This work was supported by Deutsche Telekom Laboratories and in part by the Boeing Company and the Vietnam Education Foundation.}
\thanks{Tansu Alpcan is with Deutsche Telekom Laboratories, Ernst-Reuter-Platz 7, D-10587 Berlin, Germany
        {\tt\small tansu.alpcan@telekom.de}}%
\thanks{Tamer Ba\c{s}ar and Kien C. Nguyen are with the Department of Electrical and Computer Engineering and the Coordinated Science Laboratory, University of Illinois at Urbana-Champaign, 1308 W Main St., Urbana, IL 61801, USA
        {\tt\small basar1@illinois.edu, knguyen4@illinois.edu}}%
}

\usepackage{srcltx} 
\usepackage{times,latexsym,array,verbatim,url}
\usepackage{epsfig,color,graphicx,cite,setspace}
\usepackage{mathptmx}       
\usepackage{helvet}         
\usepackage{courier}        
\usepackage{type1cm}        
\usepackage{graphicx}        
\usepackage{multicol}        
\usepackage[bottom]{footmisc}
\usepackage{psfrag}

\newtheorem {proposition} {Proposition}

\newtheorem{example}{Example}

\begin{document}

\maketitle
\thispagestyle{empty}
\pagestyle{empty}

\begin{abstract}

This paper studies a stochastic game theoretic approach to security and intrusion detection in communication and computer networks. Specifically, an \textit{Attacker} and a \textit{Defender} take part in a two-player game over a network of nodes whose security assets and vulnerabilities are correlated. Such a network can be modeled using weighted directed graphs with the edges representing the influence among the nodes. The game can be formulated as a non-cooperative zero-sum or nonzero-sum stochastic game. However, due to correlation among the nodes, if some nodes are compromised, the effective security assets and vulnerabilities of the remaining ones will not stay the same in general, which leads to complex system dynamics. We examine existence, uniqueness, and structure of the solution and also provide numerical examples to illustrate our model.

\end{abstract}

\section{INTRODUCTION} \label{sec:intro}

Today, as computer networks become ubiquitous, network security and \textit{intrusion detection} (ID) play a more and more important role. The main task of an \textit{intrusion detection system} (IDS) is to detect intrusions and report them to a system administrator. Among various approaches, non-cooperative game theory has recently been employed extensively to study ID problems \cite{ABC03, ABC04, ABI, LCMGN, Chen, Sall}.

In a general setting, a security game is defined between two players: an Attacker and a Defender (the IDS). A formulation of security games as static games can be found in \cite{ABC03}. In \cite{ABI}, the authors consider security games with imperfect observations and use the finite-state Markov chain framework to analyze such games. The work in \cite{LCMGN} employs the framework of Bayesian games to address the intrusion detection problem in wireless ad hoc networks, where a mobile node viewed as a player confronts an opponent whose \emph{type} is unknown. 

In \cite{Chen}, the author examines the intrusion detection problem in heterogenous networks as a nonzero-sum static game. In a complex network, nodes are of different levels of importance to the Defender, and also appear variably attractive to the Attacker. Heterogeneity also stems from hierarchy and correlation among nodes. It is thus essential to consider scenarios where nodes have different security assets. Also, apart from a node's security asset, if we take into account the players' motivations, the cost of attacking, the cost of monitoring, and other factors, the game is no longer a zero-sum one. Using the \textit{Nash Equilibrium} (NE) solution concept, the analysis allows one to compute the Attacker's optimal strategy as a  probability mass distribution on the nodes to attack. Similarly, the Defender's optimal strategy is a probability mass distribution on the nodes to monitor (to collect and process data and detect attacks). However, in this work \cite{Chen}, the security assets are still assumed to be independent. Also, the dynamics of the ID problem when nodes are compromised along the play have not been taken into account.

The work in \cite{Sall} addresses this problem using the framework of zero-sum stochastic games \cite{Owen}. The network is now modeled as a discrete-time or continuous-time Markov chain where the network states are defined by the states (compromised or not) of the constituent nodes. This formulation thus takes into account the dynamics of the problem and allows one to incorporate correlation among nodes in terms of vulnerability. The analysis is nonetheless limited to zero-sum games and again, the security assets are considered to be independent.

This paper attempts to extend these earlier works to construct a more comprehensive network security and intrusion detection model. We develop a network model based on \textit{linear influence networks} proposed in \cite{MK}. This model, when used under the framework of stochastic games, permits us to take into consideration the correlation among the nodes in terms of both security assets and vulnerabilities. 

The rest of this paper is organized as follows. In the remaining part of this section, we summarize the notations and variables used throughout this paper. Next, in Section \ref{sec:linear_inf_net}, we introduce two linear influence network models for security assets and vulnerabilities. In Section \ref{sec:zs_comp_games}, we formulate the security game based on these models as a zero-sum stochastic game and present results on existence, uniqueness, and structure of the solution. We then provide a numerical example in Section \ref{sec:numex}. Finally, some concluding remarks of Section \ref{sec:conclusion} end the paper.

\subsection*{Summary of notations and variables used in this paper}

\begin{itemize}
\item
$\mathcal{N}$: Set of nodes in the network.
\item
$n$: Number of nodes in the network.
\item
$\mathcal{E}_s$: Set of edges representing the influence among node security assets.
\item
$\mathcal{E}_v$: Set of edges representing the influence among node vulnerabilities.
\item
$e_{ij}$: A directed edge from node $i$ to node $j$, $e_{ij} \in \mathcal{E}_s$ or $e_{ij} \in \mathcal{E}_v$.
\item
$\mathcal{G}_s$: Weighted directed graph for node security assets, $\mathcal{G}_s= \{ \mathcal{N}, \mathcal{E}_s \}$
\item
$\mathcal{G}_v$: Weighted directed graph for node vulnerabilities, $\mathcal{G}_v= \{ \mathcal{N}, \mathcal{E}_v \}$
\item
$I,\ I_{ij}$: Influence matrix for security assets and its entries.
\item
$w_{ij}$: Influence of node $i$ on node $j$ in terms of security assets, where $i,j \in \mathcal{N}$
\item
$s=\{ s_1, s_2, \ldots, s_n \}$: Vector of independent security assets. 
\item
$x=\{ x_1, x_2, \ldots, x_n \}$: Vector of effective security assets.
\item
$H, \ h_{ij}$: Support matrix and its entries, $h_{ij}$ signifies the support that node $i$ gives node $j$ (against attacks), $0 \leq h_{ij} \leq 1 \ \forall i,j \in \mathcal{N}$. 
\item
$h_{j}$: Support to node $j, \ j \in \mathcal{N}$, $h_{j} = \sum_{i=1}^n h_{ij}$.
\item 
$p^j_{n1}$: Probability that node $j$ is compromised when player $1$ (the Attacker) attacks, player $2$ (the Defender) does not defend the node, and the support to node $j$ is equal to $1$ (full support).
\item
$p^j_{n0}$: Probability that node $j$ is compromised when the Attacker attacks, the Defender does not defend the node, and the support to node $j$ is equal to $0$ (no support).
\item
$p^j_{d1}$: Probability that node $j$ is compromised when the Attacker attacks, the Defender defends the node, and the support to node $j$ is equal to $1$ (full support).
\item
$p^j_{d0}$: Probability that node $j$ is compromised when the Attacker attacks, the Defender defends the node, and the support to node $j$ is equal to $0$ (no support).
\item
$\left\{ S_1,S_2,\ldots S_p \right\}  \ $: States in the state space of the system.
\item
$\left\{ \Gamma_1,\Gamma_2,\ldots \Gamma_p \right\}$: Game elements of the stochastic game, each of which corresponds to a state of the system.
\item
$p^k_r$: Probability that the network goes back to state $S_1$, given that it is currently in state $S_k$, the Attacker attacks
one node and the attack fails.
\item
$p^k_e$: Probability that the game ends given that it is currently in state $S_k$, the Attacker attacks one node and the attack fails.
\item
$p^k_{\emptyset r}$: Probability that the network goes back to state $S_1$, given that it is currently in state $S_k$ and the Attacker does not attack any node.
\item
$p^k_{\emptyset e}$: Probability that the game ends given that it is currently in state $S_k$ and the Attacker does not attack any node.
\item
$a^k_{ij}$: Instant amount that player $2$ pays player $1$ at game element $\Gamma_k$, if player $1$ plays pure strategy $i$ and player $2$ plays pure strategy $j$.
\item
$q_{ij}^{kl}$: Probability that both players have to play game element $\Gamma_l$ next, given that they are currently at game element $\Gamma_k$, if player $1$ plays pure strategy $i$ and player $2$ plays pure strategy $j$.
\item
$q_{ij}^{k0}$: Probability that the game ends given that they are currently at game element $\Gamma_k$, if player $1$ plays pure strategy $i$ and player $2$ plays pure strategy $j$.
\item
$m_k$: Number of pure strategies for player $1$ at game element $\Gamma_k$.
\item
$n_k$: Number of pure strategies for player $2$ at game element $\Gamma_k$.
\item
$p \ (p=2^n)$: Number of game elements of the stochastic game, or the number of states of the state space.
\item
$\alpha^k_{ij}$: A collective entry that includes the instant payoff and the transition probabilities to all game elements, $\alpha^k_{ij} = a^k_{ij} + \sum_{l=1}^p q_{ij}^{kl} \Gamma_l$, given that the players are currently at game element $\Gamma_k$, player $1$ plays pure strategy $i$, and player $2$ plays pure strategy $j$.
\item
$b^k_{ij}$: Value of $\alpha^k_{ij}$ when we replace game elements $\Gamma_l$'s with their values. $b^k_{ij} = a^k_{ij} + \sum_{l=1}^p q_{ij}^{kl} v_l$.
\item
$y_{i}^{kt}$: Probability that player $1$ plays pure strategy $i$ when playing game element $\Gamma_k$ at the $t$-th stage of the game. For stationary strategies \cite{Owen}, the superscript $t$ will be omitted.
\item
$z_j^{kt}$: Probability that player $2$ plays pure strategy $j$ when playing game element $\Gamma_k$ at the $t$-th stage of the game.
\item
$y^{kt}, \ (k=1, \ldots,p, \ t=1,2,\ldots)$: Strategy for player $1$, a set of $m_k$-vectors each of which is a mixed strategy of player $1$ at game element $\Gamma_k$ and $t$-th stage of the game.
\item
$z^{kt}, \ (k=1, \ldots,p, \ t=1,2,\ldots)$: Strategy for player $2$, a set of $n_k$-vectors each of which is a mixed strategy of player $2$ at game element $\Gamma_k$ and $t$-th stage of the game.
\item
$c^k_i$: Pure strategy $i$ for the Attacker at game element $\Gamma_k$.
\item
$d^k_j$: Pure strategy $j$ for the Defender at game element $\Gamma_k$.
\item
$p^k_{s}(c^k_i,d^k_j)$: Probability that the attack is successful given that the Attacker plays pure strategy $c^k_i$ and the Defender plays pure strategy $d^k_j$ at game element $\Gamma_k$.
\item
$v = (v_1, v_2, \ldots, v_p)$: Value vector of the stochastic game.
\item
$val(B)$: Value of the zero-sum matrix game given by the matrix $B$.

\end{itemize}

\section{LINEAR INFLUENCE NETWORK MODELS FOR SECURITY ASSETS AND FOR VULNERABILITIES} \label{sec:linear_inf_net}
We present in this section a network model based on the concept of linear influence networks \cite{MK}. The network will be represented by two weighted directed graphs, one signifying the relationship of security assets and the other denoting vulnerability correlation among the nodes.
\subsection{Linear influence network model for security assets}
For a particular node, the general term \textit{security asset} is used to signify how important the node is to the network. All the security assets of a network can be modeled as a weighted directed graph $\mathcal{G}_s= \{ \mathcal{N}, \mathcal{E}_s \}$ where $\mathcal{N}$ is the set of nodes, and the elements of set $\mathcal{E}_s$ represent the influence among the nodes. Let $n$ be the cardinality of $\mathcal{N}$. 
For each edge $e_{ij} \in \mathcal{E}_s$, we denote an associated scalar $w_{ij}$ that signifies the influence of node $i$ on node $j$, where $i,j \in \mathcal{N}$. The entries of the \textit{influence matrix} $I$ are then given as follows:
\begin{equation}
I_{ij} = \left\{
		\begin{array}{ccc}
			&w_{ij}& \textrm{ if } e_{ij} \in \mathcal{E}_s  \\
			&0& \textrm{ otherwise,}
		\end{array}
		\right.
\end{equation}
where $0 < w_{ij} \leq 1 \ \forall i,j \in \mathcal{N}$ and $\sum_{i=1}^n w_{ij} = 1,\ \forall j \in \mathcal{N}$. Note that here we allow for	the edges of the form $w_{jj} = 1-\sum_{i=1,i \neq j}^{n} w_{ij}$, which signifies the portion of influence of a node on the independent security asset of itself.

Let $s=\{ s_1, s_2, \ldots, s_n \}$ be the vector of \textit{independent security assets}. The vector of \textit{effective security assets}, denoted by $x=\{ x_1, x_2, \ldots, x_n \}$ can then be computed by the \textit{influence equation}:
\begin{equation} \label{eq:influence}
x = Is.
\end{equation}

With the condition $\sum_{i=1}^n w_{ij} = 1, \forall j=  \in \mathcal{N}$, we have that
\begin{eqnarray}
\nonumber \sum_{i=1}^n x_i &=& \sum_{i=1}^n \sum_{j=1}^n w_{ij} s_j = \sum_{j=1}^n \sum_{i=1}^n  w_{ij} s_j \\
&=& \sum_{j=1}^n s_j \sum_{i=1}^n  w_{ij} = \sum_{j=1}^n s_j.
\end{eqnarray}

Therefore, the sum of all the effective security assets is equal to the sum of all the independent security assets. The influence matrix thus signifies the redistribution of security assets. The independent security asset of a node $i$ is redistributed to all the nodes in the network that have influence on $i$ (including itself). When a node is down, the node itself and all the edges connected to it will be removed from the graph. Thus the security loss of the network will be the node's effective security asset (instead of its independent security asset). Conversely, if a node is brought back to the network, it regains its original influence on other nodes. In either case, the entries of the influence matrix have to be normalized to satisfy $\sum_{i=1}^n w_{ij} = 1,\ \forall j \in \mathcal{N}$. For a quick justification of this linear influence model, consider a GSM network, where a base station controller (BSC) $i$ controls several base transceiver stations (BTS), including BTS $j$. If a BSC fails, all the BTSs connected to it will be out of service. On the contrary, if only one BTS is compromised, the communication among the subscribers under other BTSs should not be affected (provided that the rest of the network is up and running). In such a situation, we can have for example, $w_{jj}=0.7$ and $w_{ij}=0.3$. If the BSC is down, there is still an amount of security asset $0.7s_j$ left, even though the BTS is not in service anymore. The reason is that, if this BTS gets connected to another BSC (or if the original BSC is up again), they will together create an added security asset for the network. We present in what follows an example to illustrate the linear influence network model.
\begin{example} \label{ex:network_down}
\begin{psfrags}
\psfrag{1}{$1$}
\psfrag{2}{$2$}
\psfrag{3}{$3$}
\psfrag{1}{$1$}
\psfrag{2}{$2$}
\psfrag{3}{$3$}
\psfrag{0.9}{$0.9$}
\psfrag{0.2}{$0.2$}
\psfrag{0.1}{$0.1$}
\psfrag{0.7}{$0.7$}
\psfrag{1/8}{$1/8$}
\psfrag{7/8}{$7/8$}
\psfrag{w32}{$w_{32}$}
\psfrag{w12}{$w_{12}$}
\psfrag{w31}{$w_{31}$}
\psfrag{w33}{$w_{33}$}
\psfrag{w22}{$w_{22}$}
\psfrag{w11}{$w_{11}$}
\begin{figure}[h]
  \centering
  \includegraphics[width=4cm]{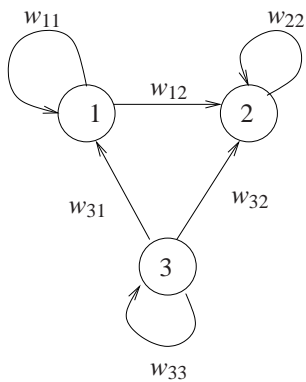}
  \caption{A linear influence network for security assets of a three-node network.} \label{fig:network3n}
\end{figure}
\end{psfrags}
\begin{psfrags}
\psfrag{1}{$1$}
\psfrag{2}{$2$}
\psfrag{3}{$3$}
\psfrag{0.9}{$0.9$}
\psfrag{0.2}{$0.2$}
\psfrag{0.1}{$0.1$}
\psfrag{0.7}{$0.7$}
\psfrag{1/8}{$1/8$}
\psfrag{7/8}{$7/8$}
\psfrag{S1}{$S_1$}
\psfrag{S2}{$S_2$}
\psfrag{S3}{$S_3$}
\psfrag{S4}{$S_4$}
\psfrag{S5}{$S_5$}
\psfrag{S6}{$S_6$}
\psfrag{S7}{$S_7$}
\psfrag{S8}{$S_8$}

Suppose that we have a network of three nodes with correlations as shown in Fig.  \ref{fig:network3n}. As shown in Fig.	\ref{fig:network3nsp}, the states of the system are given as $\left\{ S_1,S_2,\ldots S_p \right\}  \ (p=2^n)$ where $S_k \in \left\{0,1 \right\}^n, \ k=1,\ldots,p$. Here a node is said to be in state $1$ if it is compromised and $0$ otherwise. Note that we consider a discrete-time Markov chain where the system can transit from one state to any state of the state space (including the original state).
\begin{figure}[h]
  \centering
  \includegraphics[width=8.5cm]{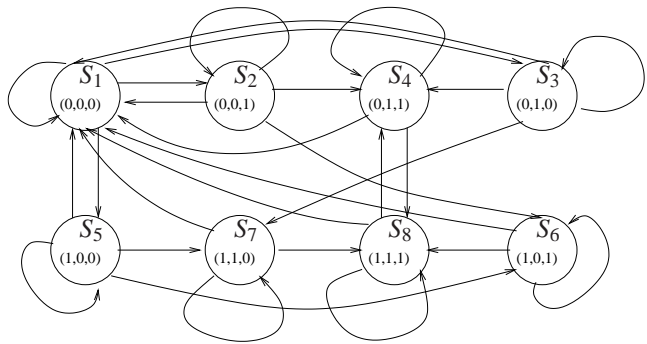}
  \caption{An example state diagram for the network in Fig. \ref{fig:network3n}.} \label{fig:network3nsp}
\end{figure}
\end{psfrags}
\begin{psfrags}
\psfrag{1}{$1$}
\psfrag{2}{$2$}
\psfrag{3}{$3$}
\psfrag{0.9}{$0.9$}
\psfrag{0.2}{$0.2$}
\psfrag{0.1}{$0.1$}
\psfrag{0.7}{$0.7$}
\psfrag{1/8}{$1/8$}
\psfrag{7/8}{$7/8$}
\begin{figure}[h]
  \centering
  \includegraphics[width=8cm]{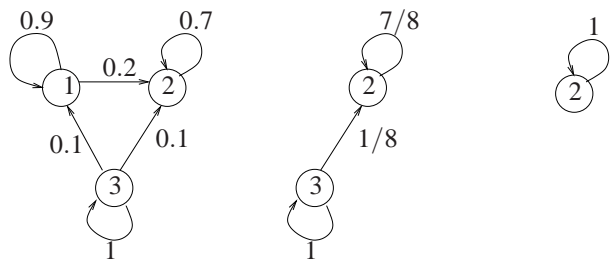}
  \caption{Changes in a linear influence network for security assets when nodes are compromised (Example \ref{ex:network_down}).} \label{fig:network3n1d}
\end{figure}
\end{psfrags}
The influence equation (\ref{eq:influence}) can be written as:
\begin{equation}
\left(
\begin{array}{c}
	x^{(1)}_1 \\
	x^{(1)}_2 \\
	x^{(1)}_3 \\
\end{array}
\right) = \left(
\begin{array}{ccc}
	0.9	&	0.2 & 0 \\
	0 & 0.7 & 0 \\
	0.1 & 0.1 & 1 \\
\end{array}
\right)
\left(
\begin{array}{c}
	s^{(1)}_1 \\
	s^{(1)}_2 \\
	s^{(1)}_3 \\
\end{array}
\right)
\end{equation}

Now suppose that node $1$ is compromised; then the independent security asset of node $3$ will remain the same, $s^{(2)}_3=s^{(1)}_3$. The independent security asset of node $2$ will be decreased by an amount corresponding to the influence of node $1$ on node $2$: $s^{(2)}_2=s^{(1)}_2-0.2s^{(1)}_2 = 0.8s^{(1)}_2$. Also, the influences on each node have to be normalized to have $\sum_{i} w_{ij} = 1$. Thus we now have $w_{32}=1/8$ and $w_{22}=7/8$, and the influence equation becomes
\begin{equation}
\left(
\begin{array}{c}
	x^{(2)}_2 \\
	x^{(2)}_3 \\
\end{array}
\right) = \left(
\begin{array}{ccc}
	7/8	&	0  \\
	1/8 & 1 \\
\end{array}
\right)
\left(
\begin{array}{c}
	s^{(2)}_2 \\
	s^{(2)}_3 \\
\end{array}
\right)
\end{equation}
Thus we can see 
\begin{eqnarray}
\nonumber x^{(2)}_2 &=& (7/8) s^{(2)}_2 = 0.7s^{(1)}_2, \\
\nonumber x^{(2)}_3 &=& (1/8) s^{(2)}_2+s^{(2)}_3 = 0.1s^{(1)}_2+s^{(1)}_3. 
\end{eqnarray}

After node $1$ goes down, the effective security asset of node $2$ remains the same, while that of node $3$ is decreased by an amount representing its influence on node $1$.

Now if node $3$ is in turn compromised, we have a network with one node as in Fig. \ref{fig:network3n1d}. We have
\begin{eqnarray}
\nonumber  s^{(3)}_2 &=& s^{(2)}_2-s^{(2)}_2/8 = (7/8)s^{(2)}_2=0.7s^{(1)}_2,\\
\nonumber  x^{(3)}_2 &=& s^{(3)}_2.
\end{eqnarray} 
\end{example}

\subsection{Linear influence network model for vulnerabilities} \label{sub:linnetvul}
In this subsection, we use the linear influence network model to represent the correlation of node vulnerabilities in a network. Beside the correlation of security assets, nodes also have influence on others' vulnerabilities. For example, within a corporate network, if a workstation is compromised, the data stored in this computer can be exploited in attacks against other workstations; these latter computers thus will become more vulnerable to intrusion. Under the framework of stochastic games, this kind of influence is readily incorporated. For instance, in the network of Example \ref{ex:network_down}, if the Attacker attacks node $1$, and the Defender decides not to defend this node, the probability that the system goes from $(0, 1, 0)$ to $(1,1,0)$ will be greater that the probability that the system goes from $(0, 0, 0)$ to $(1,0,0)$, if node $2$ has some influence on node $1$ in terms of vulnerability. For $e_{ij} \in \mathcal{E}_v$, we define the \textit{support matrix} as follows
\begin{equation}
H = \left\{
		\begin{array}{ccc}
			&h_{ij}& \textrm{ if } e_{ij} \in \mathcal{E}_v  \\
			&0& \textrm{ otherwise,}
		\end{array}
		\right.
\end{equation}
where $h_{ij}$ signifies the support that node $i$ gives node $j$ (against attacks), $0 \leq h_{ij} \leq 1 \ \forall i,j \in \mathcal{N}$. The \textit{support} to node $j, \ j \in \mathcal{N}$ is defined as
\begin{equation}
h_{j} = \sum_{i=1}^n h_{ij},
\end{equation} 
where $0 \leq h_j \leq 1,\ \forall j \in \mathcal{N}$. Unlike the model for security assets, here we do not normalize $h_j$. When a node that supports node $j$ is down, $h_j$ will decrease, and thus the probability that node $j$ is compromised under attack will increase. Let us denote by $p^j_s$ the probability that node $j$ is compromised at each state. We assume an affine relationship between $p^j_s$ and $h_j$ as follows:
\begin{itemize}
	\item 
	If node $j$ is not attacked then $p^j_s=0$.
	\item
	If node $j$ is attacked, and the Defender is not defending this node, $p^j_s = p^j_{n0} - (p^j_{n0}-p^j_{n1})h_j$, where $p^j_{n1}$ and $p^j_{n0}$ are the probabilities that the node is compromised given that the support is equal to $1$ (full support) and $0$ (no support), respectively ($p^j_{n1}<p^j_{n0}$).
	\item
	If node $j$ is attacked, and the Defender is defending this node, $p^j_s = p^j_{d0} - (p^j_{d0}-p^j_{d1})h_j$, where $p^j_{d1}$ and $p^j_{d0}$ are the probabilities that the node is compromised given that the support is equal to $1$ and $0$, respectively ($p^j_{d1}<p^j_{d0}$).
	\item
	Also, it is assumed that $p^j_{d1}< p^j_{n1}$ and $p^j_{d0}< p^j_{n0}$.
\end{itemize}
A weighted directed graph for network vulnerabilities is shown in Fig. \ref{fig:netvul}.
\begin{psfrags}
\psfrag{1}{$1$}
\psfrag{2}{$2$}
\psfrag{3}{$3$}
\psfrag{0.9}{$0.9$}
\psfrag{0.2}{$0.2$}
\psfrag{0.1}{$0.1$}
\psfrag{0.7}{$0.7$}
\psfrag{0.5}{$0.5$}
\psfrag{0.3}{$0.3$}
\psfrag{1/8}{$1/8$}
\psfrag{7/8}{$7/8$}
\begin{figure}[h]
  \centering
  \includegraphics[width=8cm]{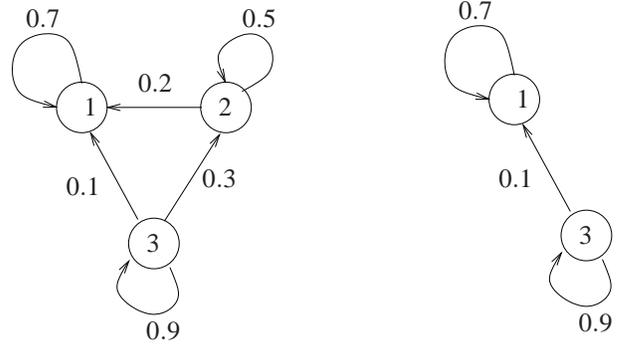}
  \caption{A linear influence network for vulnerabilities and the changes of supports when one node is compromised.} \label{fig:netvul}
\end{figure}
\end{psfrags}
\section{THE NETWORK SECURITY PROBLEM AS A ZERO-SUM STOCHASTIC GAME}  \label{sec:zs_comp_games}
\subsection{A brief overview of zero-sum stochastic games} \label{sub:overview_zs}
In this subsection, we provide a brief overview of zero-sum stochastic games based on \cite{Owen}. A stochastic game consists of $p$ game elements $\Gamma_k, \ k=1,\ldots,p$. Each game element is associated with an $m_k \times n_k$ matrix, whose entries are given by
\begin{equation} \label{sg_entries}
\alpha^k_{ij} = a^k_{ij} + \sum_{l=1}^p q_{ij}^{kl} \Gamma_l,
\end{equation}
\begin{eqnarray}
\nonumber \textrm{where   } q_{ij}^{kl} &\geq& 0,\ l = 1,\ldots,p,\ i = 1,\ldots,m_k,\  j = 1,\ldots,n_k,\\
\label{game_end} \sum_{l=1}^p q_{ij}^{kl} &<& 1, \ \forall k, i,j.
\end{eqnarray}
Expression (\ref{sg_entries}) can be interpreted as follows. At game element $\Gamma_k$, if player $1$ chooses pure strategy $i$ and player $2$ chooses pure strategy $j$, player $2$ has to pay player $1$ an amount $a^k_{ij}$. Furthermore, there is a probability $q_{ij}^{kl}$ that both players have to play game element $\Gamma_l$ next, and a probability
\begin{equation}
q_{ij}^{k0} = 1 - \sum_{l=1}^p q_{ij}^{kl} 
\end{equation}
that the game will end. With condition (\ref{game_end}), the probability of infinite play is guaranteed to be zero, and the expected payoff of player $1$ (or the expected loss of player $2$), which is accumulated through all the stages of the game, is finite \cite{Owen}.

A strategy for player $1$ is a set of $m_k$-vectors, denoted by $y^{kt}, \ k=1,\ldots,p, \ t=1,2,\ldots$, each of which satisfies
\begin{eqnarray}
\sum_{i=1}^{m_k} y_{i}^{kt} =1, \\
y_{i}^{kt} \geq 0
\end{eqnarray}
Here $y_{i}^{kt}$ is the probability that player $1$ plays pure strategy $i$ if he is playing game element $\Gamma_k$ at the $t$-th stage of the game. A strategy is said to be stationary if the vectors $y^{kt}$ are independent of $t$ for all $k$. In this case, the superscript $t$ can be omitted. Similarly, a strategy for player $2$ is a set of $n_k$-vectors, $z^{kt}$, where $\sum_{j=1}^{n_k} z_j^{kt} =1$ and $z_j^{kt} \geq 0$. Given a pair of strategies, we can compute the vector of expected payoffs $v = (v_1, v_2, \ldots, v_p)$, where $v_k,\ k=1,\ldots,p$ is the expected payoff (to player $1$) if the first stage of the game is $\Gamma_k$. 

With the above settings, it is known \cite{Owen}, that we can replace the game element $\Gamma_k$ by the value component
\begin{equation} \label{val}
v_k = val(B_k),
\end{equation}
where $val(B_k)$ is the value (in mixed strategies) of the matrix game $B_k$, and $B_k$ is the $m_k \times n_k$ matrix whose entries are given by
\begin{equation} \label{b_k}
b^k_{ij} = a^k_{ij} + \sum_{l=1}^p q_{ij}^{kl} v_l.
\end{equation}
\subsection{A zero-sum stochastic game model for network security} \label{sub:zsgame}
In this subsection we formulate the security problem as a zero-sum stochastic game. This is a modified version of the game presented in \cite{Sall}, applied to the linear influence network model proposed in Section \ref{sec:linear_inf_net}. 	
At each state $k,\ k = 1, \ldots, p$, the Attacker's pure strategies consist of $m_k=n+1$ actions, where $n$ is the number of nodes in the network:
  \begin{itemize}
  \item
  Attack one of $n$ nodes, $c^k_i$, where $i=1,\ldots, n$.
  \item
  Do nothing, $c^k_{m_k}=\emptyset$.
  \end{itemize}
  Note that this strategy space is for use with more general payoff formulations. However, with the payoff formulation in this paper, the Attacker will not have motivation to attack a node that is already compromised, unless all the nodes have been compromised. For each $k$, the Defender's pure strategies are $\left\{ d^k_i \right\}$, where 
  
\begin{itemize}
	\item 
Defend node $i$, $d^k_i,i=1,\ldots, n_k-1$, 
\item
Do nothing, $d^k_{n_k}=\emptyset$, 
\end{itemize}
where $n_k=m_k=n+1$. For each possible combination of the Attacker's and the Defender's pure strategies, the entries of the payoff matrix are:
  \begin{eqnarray}
	\alpha^k_{ij} = a^k_{ij} + \sum_{l=1}^p q_{ij}^{kl} \Gamma_l,
  \end{eqnarray}
	where $a^k_{ij}=p^k_{s}(c^k_i,d^k_j)x^k(i)$, $p^k_{s}(c^k_i,d^k_j)$ is the probability that the attack is successful, and $x^k(i)$ is the effective security asset of the node being attacked, $i$. Note that once a node is compromised, the effective security assets and the supports of the remaining nodes have to be recalculated as in Example \ref{ex:network_down} and Fig.  \ref{fig:netvul}. As mentioned in Subsection  \ref{sub:linnetvul}, the probabilities $p^k_{s}$, and thus $q_{ij}^{kl}$, are dependent on the supports to the nodes, and are therefore affected by the correlation in vulnerabilities of the nodes. It can be said that once we have incorporated node vulnerabilities into our model, we have already implicitly taken care of the cost of attacking/defending. For example, if a node is of high security asset but difficult to compromise (the transition probability to the compromise state is small), the Attacker may turn to another node with a smaller security asset, which is easier to attack.	
	
	At a state $S_k$, if the Attacker chooses to attack one node and the attack fails, there is a probability $p^k_r \in (0,1)$ that the network will go back to state $S_1$ (which means the Defender has detected the Attacker and managed to restore all the compromised nodes and the game restarts at $S_1$), and a probability $p^k_e \in (0,1)$ that the game will end (which means the Defender has detected the Attacker and stopped him from further intruding). Note that $p^k_r+p^k_e \leq 1$ with equality only when $S_k = S_1 (0,0,\ldots,0)$. Similarly, at one point, if the Attacker chooses not to attack at all, there is a probability $p^k_{\emptyset r} \in (0,1)$ that the network will go back to state $S_1$, and a probability $p^k_{\emptyset e} \in (0,1)$ that the game will end.
	Given $0<p^j_{d1},\ p^j_{n1},\ p^j_{d0},\ p^j_{n0}<1, \ j \in \mathcal{N}, \ p^k_r$, \ $p^k_e, \ p^k_{\emptyset r}$, and $p^k_{\emptyset e}, \ k = 1,\ldots,p$, and the support matrix $H$, $p^k_{s}$ and $q_{ij}^{kl}$ can be calculated using the equations in Subsection \ref{sub:linnetvul}. A numerical example is shown in Section \ref{sec:numex}.
	
\subsection{Existence, uniqueness, and structure of the solution} \label{sub:zs_comp_games}
We present in this subsection some analytical results for the game given in \ref{sub:zsgame}, based on zero-sum stochastic game theory \cite{Owen}, \cite{Shapley}.
\begin{proposition}
In the zero-sum stochastic game given in \ref{sub:zsgame}, the probability of infinite play is zero and the expected payoff of the Attacker (which is also the expected cost of the Defender) is finite.
\end{proposition}
With the setup in \ref{sub:zsgame}, we can show that $q_{ij}^{k0} = 1 - \sum_{l=1}^p q_{ij}^{kl}>0, \ \forall k$ and $\forall \ i,j$ of each game element $\Gamma_k$. Thus the proposition is proved using the theory of stochastic games.
\begin{proposition} (Theorem $V.3.3$ \cite{Owen})
In the zero-sum stochastic game given in \ref{sub:zsgame}, there exists exactly one vector $v = (v_1, v_2, \ldots, v_p)$ that satisfies (\ref{val}) and (\ref{b_k}).
\end{proposition}
Using the results from  \ref{sub:overview_zs}, we can then compute the NE of the game, which is a pair of stationary mixed strategies for the Attacker and for the Defender at each state. 
\begin{proposition} (Theorem $V.3.3$ \cite{Owen}) \label{recproc}
The vector $v = (v_1, v_2, \ldots, v_p)$ that satisfies (\ref{val}) and (\ref{b_k}) can be derived through the following recursive equations:
\begin{eqnarray} \label{zs_rec}
v^0 &=& (0, 0, \ldots, 0), \\
b^{kr}_{ij} &=& a^k_{ij} + \sum_{l=1}^p q_{ij}^{kl} v^r_l, \\
v^{r+1}_{k} &=& val(B^r_k) = val(b^{kr}_{ij}).
\end{eqnarray}
\end{proposition}
We can stop the recursion at a desired level of accuracy and then use the current value of vector $v = (v_1, v_2, \ldots, v_p)$ to compute $B_k$ using (\ref{b_k}). The mixed strategies of the players at each game element $\Gamma_k$ are the NE in mixed strategies of the matrix game $B_k$.  The strategies so obtained will converge to optimal stationary strategies of the stochastic game.
\section{A NUMERICAL EXAMPLE} \label{sec:numex}
In this section, we implement numerical simulation for a specific network with three nodes. The setup in \ref{sub:zsgame} is carried over with some further assumptions as follows. First, we adopt a simplified state diagram as given in Fig. \ref{ex:network_down}. Basically, after each time step, we only allow for transitions where one more node is compromised, the transition that returns to the same state, and the transition back to $S_1 (0,0,0)$. Second, suppose that the influence equation is given as follows (Example \ref{ex:network_down})
\begin{equation}
\left(
\begin{array}{c}
	x^{(1)}_1 \\
	x^{(1)}_2 \\
	x^{(1)}_3 \\
\end{array}
\right) =
\left(
\begin{array}{ccc}
	0.9	&	0.2 & 0 \\
	0 & 0.7 & 0 \\
	0.1 & 0.1 & 1 \\
\end{array}
\right)  \left(
\begin{array}{c}
	10 \\
	10 \\
	20 \\
\end{array}
\right)=\left(
\begin{array}{c}
	11 \\
	7 \\
	22 \\
\end{array}
\right),
\end{equation}
and the support matrix is given by (Fig. \ref{fig:netvul})
\begin{equation}
H = \left(
\begin{array}{ccc}
	0.7	&	0 & 0 \\
	0.2 & 0.5 & 0 \\
	0.1 & 0.3 & 0.9 \\
\end{array}
\right).
\end{equation}
Finally, $p^j_{d1}=0.2,\ p^j_{n1}=0.4,\ p^j_{d0}=0.5,\ p^j_{n0}=0.7, \forall j \in \mathcal{N}$, $p^k_r=0.2, \ \forall k \neq 1$, $p^1_r=0.7$, $p^k_e =0.3, \ \forall k=1,\ldots,p$, $p^k_{\emptyset r}=0.2, \ \forall k \neq 1$, $p^1_{\emptyset r}=0.7$, and $p^k_{\emptyset e}=0.3, \ \forall k=1,\ldots,p$.

For example, suppose the system is at $S_1 \ (0,0,0)$. The next state could be one in $\{ S_1 \ (0,0,0), \ S_2 \ (0,0,1),  \ S_3 \ (0,1,0),  \ S_5 \ (1,0,0) \}$. The Attacker's pure strategies include $1,2,3$, and $\emptyset$, which mean to attack node $1$, node $2$, node $3$, and do nothing, respectively. Similarly, the Defender's pure strategies include $1,2,3$, and $\emptyset$.
Using the above results, we have that
\begin{eqnarray}
\nonumber a^1_{11} &=& p^1_s(1,1)x^{(1)}_1, \\ 
\nonumber q^{11}_{11} &=& (1 - p^1_s(1,1))(1-p^{1e}),\\
\nonumber q^{15}_{11} &=& p^1_s(1,1), \\ 
\nonumber q^{1j}_{11} &=& 0 \ \forall j \neq 1,5,
\end{eqnarray}
where $p^1_s(1,1)= p_{d0} - (p_{d0}-p_{d1})1 = p_{d1}$, as at this state, node $1$ still has full support. Also, there is a probability $p^{1e}_g = (1 - p^1_s(1,1))p^{1e} > 0$ that the game will end. If the Attacker attacks node $1$ and the Defender defends node $2$, we have that
\begin{eqnarray} 
\nonumber a^1_{12} &=& p^1_s(1,2)x^{(1)}_1, \\ 
\nonumber q^{11}_{12} &=& (1 - p^1_s(1,2))(1-p^{1e}), \\
\nonumber q^{15}_{12} &=& p^1_s(1,2), \\
\nonumber q^{1j}_{12} &=& 0 \ \forall j \neq 1,5,
\end{eqnarray}
where $p^1_s(1,1)= p_{n0} - (p_{n0}-p_{n1})1 = p_{n1}$, again as at this state, node $1$ still has full support. Also, there is a probability $p^{1e}_g = (1 - p^1_s(1,2))p^{1e} > 0$ that the game will end.
Now, suppose that the system is at $ S_5 \ (1,0,0)$. The next state could be one in $\{ S_1 \ (0,0,0), \ S_5 \ (1,0,0),  \ S_6 \ (1,0,1),  \ S_7 \ (1,1,0) \}$. The Attacker's pure strategies include $2,3$, and $\emptyset$, which mean to attack node $2$, node $3$, and do nothing, respectively. Similarly, the Defender's pure strategies include $2,3$, and $\emptyset$.
Now we have that
\begin{eqnarray}
\nonumber a^5_{22} &=& p^2_s(2,2)x^{(5)}_2, \\ 
\nonumber q^{57}_{22} &=& p^2_s(2,2), \\ 
\nonumber q^{51}_{22} &=& (1-p^2_s(2,2))p^5_r, \\
\nonumber q^{55}_{22} &=&  (1-p^2_s(2,2))(1-p^5_r-p^5_e), \\ 
\nonumber q^{5j}_{22} &=& 0 \ \forall j \neq 1,5,7,
\end{eqnarray}
where $p^2_s(2,2)= p^2_{d0} - (p^2_{d0}-p^2_{d1})0.8$, as at this state, node $2$ has a support of $0.8$. Also, there is a probability $p^{5e}_g = (1 - p^2_s(2,2))p^{5e} > 0$ that the game will end. The other entries of other game elements can be calculated in a similar way. Using the recursive procedure given in Proposition \ref{recproc}, we can then compute the optimal strategy of each player and the value of the game. The value vector converges to an accuracy of $10^{-4}$ after $56$ iterations. The optimal strategies of the Attacker and the Defender, and the value vector are given in Tables \ref{tab:attackerstr}, \ref{tab:defenderstr}, and 	\ref{tab:defenderstr}. As can be seen from Table \ref{tab:attackerstr}, for example, when all the nodes are up and running, the Attacker wants to attack node $1$ with probability $0.6126$ and node $3$ with probability $0.3874$, while the Defender wants to defend node $1$ with probability $0.0702$ and node $3$ with probability $0.9298$. Recall that the effective security assets of nodes $1,\ 2$, and $3$ at this state are $11,\ 7$, and $22$, respectively. It is worth noting that the mixed strategies for the players can also be interpreted as the way to allocate their resources in the security game.

\begin{table} 
	\centering
		\begin{tabular}{|l|l|l|l|l|}
		\hline
GE & Node 1 & Node 2 & Node 3 & Do nothing \\
\hline
   $1 \ (0,0,0)$ & $0.6126$ &   $0$ &    $0.3874$ &   $0$ \\
\hline    
 $2 \ (0,0,1)$ &    $0.3817$  &  $0.6183$ &   $0$ &   $0$\\
\hline    
 $3\ (0,1,0)$ &    $0.6415$   & $0$  & $0.3585$   & $0$\\
\hline    
$4 \ (0,1,1)$&    $1$     &    $0$  &  $0$ & $0$\\
\hline    
 $5 \ (1,0,0)$ &    $0$  &  $0.6568$  &  $0.3432$  & $0$\\
\hline    
 $6 \ (1,0,1)$ &    $0$  &  $1$    &     $0$   & $0$\\
\hline    
$7 \ (1,1,0)$ &    $0$   &   $0$ &   $1$  &  $0$\\
\hline         
$8 \ (1,1,1)$ &    $0.25$  &  $0.25$ &  $0.25$ &   $0.25$\\
    \hline
		\end{tabular}
	\caption{Optimal strategies for the Attacker at each game element (GE).}
	\label{tab:attackerstr}
\end{table}

\begin{table} 
	\centering
		\begin{tabular}{|l|l|l|l|l|}
		\hline
GE & Node 1 & Node 2 & Node 3 & Do nothing \\
\hline
 $1 \ (0,0,0)$ &   $0.0702$   &      $0$  &  $0.9298$   &      $0$\\
\hline    
 $2 \ (0,0,1)$ &  $0.6614$  &  $0.3386$     &    $0$    &     $0$\\
\hline
 $3\ (0,1,0)$ &  $0.0869$     &    $0$   & $0.9131$      &   $0$\\
\hline
 $4 \ (0,1,1)$ & $1$   & $0$   & $0$   & $0$\\
\hline
 $5 \ (1,0,0)$ &     $0$  &  $0.034$   & $0.966$      &   $0$\\
\hline
 $6 \ (1,0,1)$  &        $0$  &  $1$ &  $0$  & $0$\\
\hline
$7 \ (1,1,0)$ &   $0$ &  $0$  &  $1$  & $0$\\
\hline
$8 \ (1,1,1)$ &    $0.25$  &  $0.25$  &  $0.25$   & $0.25$    \\
    \hline
		\end{tabular}
	\caption{Optimal strategies for the Defender at each game element.}
	\label{tab:defenderstr}
\end{table}

\begin{table} 
	\centering
		\begin{tabular}{|l|l|l|l|l|}
		\hline
GE & $1$ & $2$ & $3$ & $4$  \\
		\hline
Payoffs & $19.6078$ &  $15.8301$ &  $17.9557$ &  $12.3392$   \\
    \hline
GE &    $5$ & $6$ & $7$ & $8$ \\
\hline
Payoffs & $17.9659$ &  $13.0283$ &  $15.3228$ &   $7.8431$ \\
\hline
		\end{tabular}
	\caption{The value vector (the expected payoffs of the Attacker, also the expected losses of the Defender at each game element).}
	\label{tab:defenderstr}
\end{table}

\section{CONCLUSION} \label{sec:conclusion}
In this paper we have proposed a new network model based on linear influence networks to represent the interdependence of nodes in terms of security assets and vulnerabilities. We took the first step to formulate the security game between an Attacker and a Defender over this network using the framework of zero-sum stochastic game theory. The optimal solution obtained allows one to comprehend the behavior of a rational attacker, as well as to provide IDSs with guidelines on how to allocate their resources. Moreover, modeling networks with linear influence network models helps facilitate solving the security games using software programs.
As mentioned earlier, apart from a node's security asset, if we take into account the players' motivations, the cost of attacking, the cost of monitoring, and other factors, the game is no longer a zero-sum one. This work thus can be extended to  nonzero-sum stochastic games, where we can address more flexible and practical payoff formulations. Furthermore, in many real-world scenarios, neither the Attacker nor the Defender has full knowledge of the network's nodes and their correlation. Thus studying stochastic security games with incomplete information is an intriguing research direction.

\section{ACKNOWLEDGMENTS}
We would like to thank Deutsche Telekom Laboratories, the Boeing Company, and the Vietnam Education Foundation for their support. We are also grateful to four anonymous reviewers for their valuable comments.

\newpage

\bibliographystyle{IEEEtran}


\end{document}